# Spin- and angle-resolved photoemission on the topological Kondo Insulator candidate: SmB$_6$

Nan Xu[1,2], Hong Ding[3,4] & Ming Shi[1]

[1] *Swiss Light Source, Paul Scherrer Institut, CH-5232 Villigen PSI, Switzerland*
[2] *Institute of Condensed Matter Physics, École Polytechnique Fédérale de Lausanne, CH-1015 Lausanne, Switzerland*
[3] *Beijing National Laboratory for Condensed Matter Physics and Institute of Physics, Chinese Academy of Sciences, Beijing 100190, China*
[4] *Collaborative Innovation Center of Quantum Matter, Beijing, China*

**Abstract**

Topological Kondo insulators are a new class of topological insulators in which metallic surface states protected by topological invariants reside in the bulk band gap at low temperatures. Unlike other three-dimensional topological insulators, a truly insulating bulk state, which is critical for potential applications in next-generation electronic devices, is guaranteed by many-body effects in the topological Kondo insulator. Furthermore, the system has strong electron correlations that can serve as a testbed for interacting topological theories. This topical review focuses on recent advances in the study of SmB$_6$, the most promising candidate for a topological Kondo insulator, from the perspective of spin- and angle-resolved photoemission spectroscopy with highlights of some important transport results.

**Email:** nan.xu@psi.ch; dingh@iphy.ac.cn; ming.shi@psi.ch.

Table of contents

# 1. Introduction.

## 1.1 Topological insulator

Quantum states of condensed matter are characterized both by energy band structure and by the nature of wave functions[1,2]. The topological property of wave functions remains unchanged under adiabatic deformations of the system so that states with different topological orders cannot be transformed into each other without forming new phases at the boundary. The properties of topological materials and of their boundaries are an increasingly important theme in modern condensed matter physics, and experimental realizations of novel topological phases have recently attracted significant interest because of their fundamental scientific importance and great potential in applications. The quantum Hall insulator (Chern insulator) was the first discovered non-trivial topological phase showing an emergent edge state. i.e., the materials that form two-dimensional (2D) electron gas exhibiting an integer quantum Hall effect (IQH)[3]. Under a strong magnetic field, the 2D electron gas can have only discrete energy values (Landau levels) due to the quantization of the cyclotron orbits. When the chemical potential sits between adjacent Landau levels, the IQH system behaves like a bulk insulator, but at the boundary incomplete cyclotron orbits lead to metallic states along the edges with quantized Hall conductance ($Ze^2/\hbar$). In IQH materials the Hall conductivity $\sigma_{xy}$ of the edge current is topologically protected and described by an integer Chern number (Thouless-Kohmoto-Nightingale-den Nijs (TKNN) integer)[4,5], which characterizes the different topological ground states.

A topological insulator (TI) is a novel quantum phase in which any boundary to an ordinary insulator always shows metallic states. The TI is classified by a two-valued topological invariant, the so-called $Z_2$ index, which is determined by the topology of the bulk wave functions. Nowadays the term "topological insulator" (coined in ref. [6]) refers in particular to this $Z_2$ TI with time-reversal symmetry (TRS). The topological invariant is a global property of a system, whereas local order parameters are embodied in Landau symmetry breaking theory[7]. A 2D TI, also known as a quantum spin Hall (QSH) insulator, was proposed in a model of graphene by Kane and Mele[8,9] in 2005, can be considered as two time-reversed copies of the quantum Hall insulator. In such topological quantum states, strong spin-orbit coupling (SOC) acts as an effective magnetic field, and therefore the pair of the IQH edge states protected by TRS can be realized without an external magnetic field. In 2006 Bernevig, Hughes and Zhang proposed HgTe/CrTe quantum wells as a candidate for the QSH insulator and established an intuitive model Hamiltonian to describe a topological phase transition driven by band inversion[10]. The QSH effect in quantum wells was observed experimentally soon after by König *et al.*[11] in 2007.

Unlike the quantum Hall insulator that is only defined for a 2D system, the concept of a TI has been extended to three-dimensional (3D) materials, in which four $Z_2$ invariants fully characterize the quantum phase. The metallic surface states of the 3D TIs protected by TRS host Dirac fermions with spin-momentum locking[9,10,12]. The peculiar spin structure of the Dirac fermions suppresses scattering by nonmagnetic impurities and disorder[13,14] and gives rise to a dissipationless spin current[15]. Various 3D TI materials have been discovered[16]. Fu and Kane predicted that the TI can be realized in $Bi_{1-x}Sb_x$ alloys, the nontrivial topology of which can be probed by mapping an odd number of Fermi surface (FS) crossings between two TR-invariant momenta[12]. This has been experimentally confirmed with angle-resolved photoemission spectroscopy (ARPES) by Hsieh *et al.*[17] Tetradymite semiconductors ($Bi_2Se_3$, $Bi_2Te_3$ and $Sb_2Te_3$) were also proposed[18] to be TIs and confirmed experimentally[19,20]. In addition, TI states are also predicted in other materials such as thallium-based III-V-VI$_2$ ternary chalcogenides[21,22], $Bi_{14}Rh_3I_9$[23], $Ag_2Te$[24], $ZrTe_5$[25], $TlN$[26], $BaBiO_3$[27], and $Li_2IrO_3$[28]. In some of these materials the TI states have been experimentally verified[23,29,30].

Besides the topological and quantum Hall insulators, there are topological materials in which the wave functions can be classified by some other topological invariants. For example, topological crystalline insulators (TCI) can be specified by a mirror Chern number invariant with a nonzero integer[31], which has been predicted[32] and experimentally realized[33,34,35,36] in SnTe. In the last decade, because of its ability to reveal electronic structure and identifying bulk and surface contributions to it, ARPES and its spin-resolved variant (SARPES) have played a vital role in the experimental confirmation of novel topological phases.

## 1.2 Topological Kondo insulator and predicted candidate: $SmB_6$

Studies of nontrivial band topology have mainly focused on weakly correlated, inversion symmetric materials such as $Sb_2Te_3$, $Bi_2Te_3$, and $Bi_2Se_3$ and have been conducted within the framework of non-interacting topological theory[37,38]. In such materials the energy gap is small and conducting channels other than the topological surface states can be easily formed[16], which poses a serious obstacle to the application of TIs in electronic devices. As a crucial way around this issue, it has been suggested that TIs with a robust insulating gap could be realized in some Kondo insulators. In a Kondo insulator, the spin orbit coupling of *f* electrons is much larger than the characteristic energy gap, and the *f* states are odd-parity while the conduction *d* states are even parity. Based on the above considerations, Dzero, Sun, Galitski and Coleman have brought forward the concept of a topological Kondo insulator (TKI) that is a strongly interacting version of the $Z_2$ topological insulator[39,40], as shown in **Fig. 1**. TIs with strong electronic correlations may host a variety of novel interaction-driven phenomena that have attracted considerable theoretical attention.

As the first candidate TKI, the mixed valence Kondo insulator $SmB_6$ is currently of high interest. In this compound the hybridization of the localized *f* electrons with conduction *d* electrons opens a narrow band gap on the order of ~ 10 meV at low temperatures, and then the chemical potential lies in the gap (see **Fig. 1c**). First principle calculations predict that $SmB_6$ has a nontrivial $Z_2$ topology and topologically protected surface states (TSS). The topologically nontrivial surface bands of $SmB_6$ are an emerging subject that helps extend our understanding of topological insulators beyond non-interacting ones. Theoretical studies offer the exciting possibility of topologically protected surface states (TSS) in strongly correlated $SmB_6$. Thus it is crucial to characterize the surface states experimentally and identify whether $SmB_6$ is the first topological Kondo insulator.

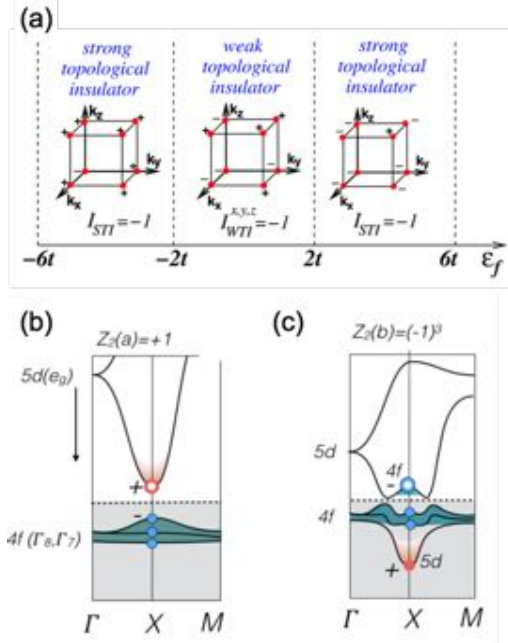

**Figure 1**. (Color online.) (a) The topological phase diagram as a function of the relative position between the *f* level and the bottom of the conduction band. (b)-(c) Band structure illustration for the *d/f* states without and with band crossing, respectively. (a) reprinted with permission from Ref. [39], and (b)-(c) from review article Ref. [41].

The transport properties of $SmB_6$ have been well studied[42,43,44]. **Figure 2** shows the resistivity of $SmB_6$ as a function of temperature (T) and 1/T. At high temperatures $SmB_6$ behaves as a correlated bad metal, and with decreasing temperature a metal-to-insulator transition occurs due to the opening of a hybridization band gap. Below ~3.5 K, however, instead of diverging as in an insulator, the resistivity saturates, indicating additional conduction channels that were suspected to originate from in-gap states – e.g., bulk impurity states[42,43,44].

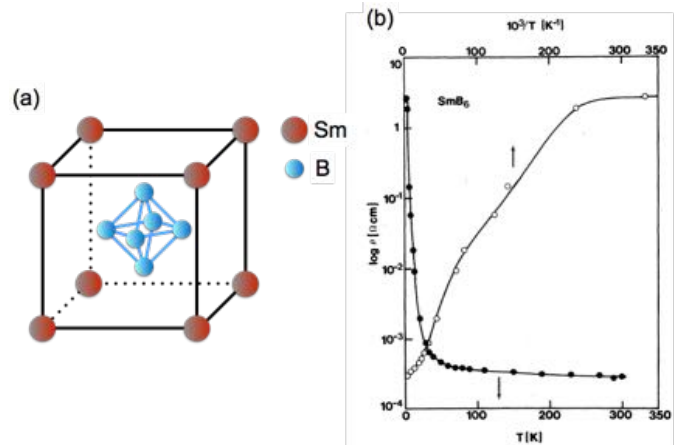

**Figure 2**: (a) Crystal structure of $SmB_6$. (b) Temperature dependence of resistivity of $SmB_6$, reproduced from ref. [42].

Recent transport experiments provided convincing evidence of surface conductivity in the perfect bulk insulator[45,46], suggesting $SmB_6$ is an ideal topological insulator for topological device applications. Point-contact spectroscopy revealed that the low-temperature Kondo insulating state harbors conduction states on the surface, in support of the nontrivial topology in the Kondo insulators [47]. However, the surface conduction obtained by transport measurements does not help elucidate the topology of these in-gap states. An ideal way to study the topology is to measure the surface states and their spin structure directly.

## 2. Electronic structure of $SmB_6$ probed by ARPES

### 2.1 Angle-resolved photoemission spectroscopy

ARPES is a modern application of the photoelectric effect[48] that directly probes the electron band structure of crystals[49,50]. By shining light on a clean metal in a vacuum vessel, electrons are emitted from the surface by the photoelectric effect. The intensity of photoelectrons at a kinetic energy ($I(E_K)$) is proportional to density of states (DOS) at a binding energy $E_B$: $I(E_K) \propto N(E_B)f(E_B)$, where $f(E_B)$ is the Fermi-Dirac distribution function. From energy conservation, the kinematic relation in the photoemission process is expressed as: $E_K = h\nu - \phi + E_B$ where $\phi$ is the work function of the sample surface. Thus photoemission spectroscopy (PES) is a powerful tool to measure the occupied DOS. In addition, band dispersions and Fermi surfaces can be determined by measuring emission angles of the photoelectrons as well as their kinetic energy.

In the photoemission process, the momentum of the photoelectron $p_{||}$ is conserved along the in-plane direction. Under the experimental configuration depicted in **Figure 3**, the wave vector of an electron inside the crystal can be expressed as:

$$\hbar k_{||} = p_{||} = \sqrt{2mE_K}\{\sin(\theta)\sin(\varphi)\vec{k}_x + \sin(\theta)\cos(\varphi)\vec{k}_y\}.$$

This relationship is valid provided that the relaxation time of the photo-holes is much longer than the escape time of the photoelectrons.

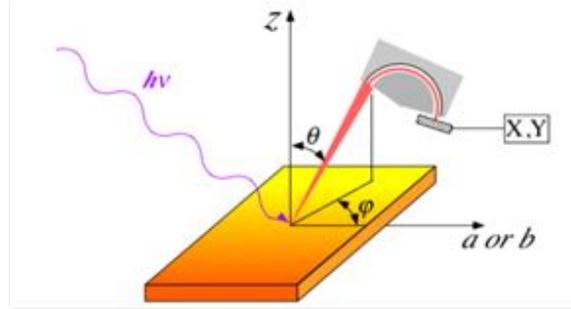

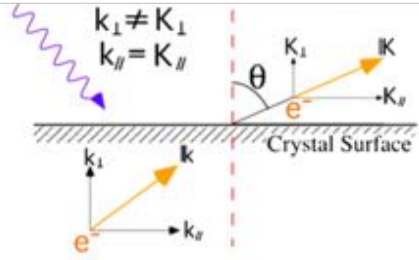

**Figure 3**. (Color online.) Schematic diagram of an angle-resolved photoemission experiment (reprinted with permission from Ref. [51]).

Although the out-of-plane wave vector ($k_\perp$) is not conserved due to the translation symmetry breaking along the normal direction at the sample surface, $k_\perp$ can be extracted by using a nearly-free electron approximation for the final states:

$$\hbar k_\perp = \sqrt{2mE_K \cos^2\theta + V_0}$$

where $V_0$ is the inner potential, which can be estimated from first-principle calculations. In modern synchrotron-based ARPES experiments, the photon energy can be tuned continuously and thus one can map out the band structure in the whole 3D momentum space.

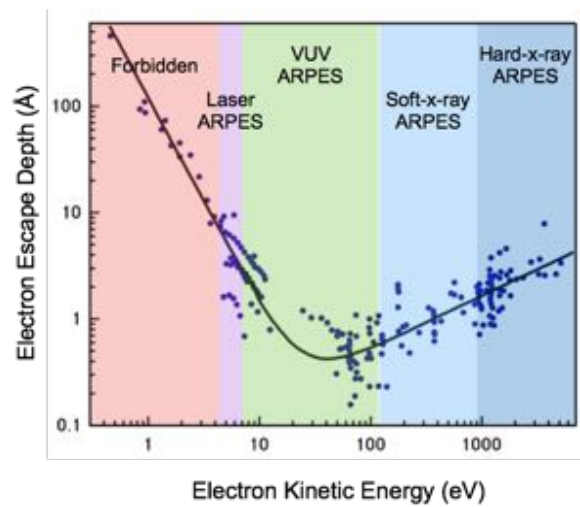

**Figure 4**. (Color online.) Universal curve of photoelectron mean free path as a function of photoelectron energy[52].

For incident photon energy in the vacuum ultraviolet (VUV) region, the small mean free path of photoelectrons with corresponding $E_K$ results in ARPES being a surface sensitive technique[52] that is powerful for revealing surface/subsurface electronic structures. As shown in **Fig. 4**, one can increase the bulk sensitivity in ARPES either by using high-energy photons in the soft X-ray region or by using low-energy photons from excitation sources. While the former is accessible with tunable synchrotron-based light[53], laser sources have the advantage of high resolution in both energy and momentum. In addition, the time structure of the light source enables ARPES with time-resolution in the pump-probe setup. By finely tuning the photon energy of the pump pulse, electrons in valence bands can be pumped into the empty conduction bands and probed by the following probing pulse (two photon photoemission). This makes the unoccupied states measurable, with a much higher cross-section and resolution than the alternative method of inverse photoemission[49].

## 2.2 Bulk hybridization gap and in-gap states in $SmB_6$

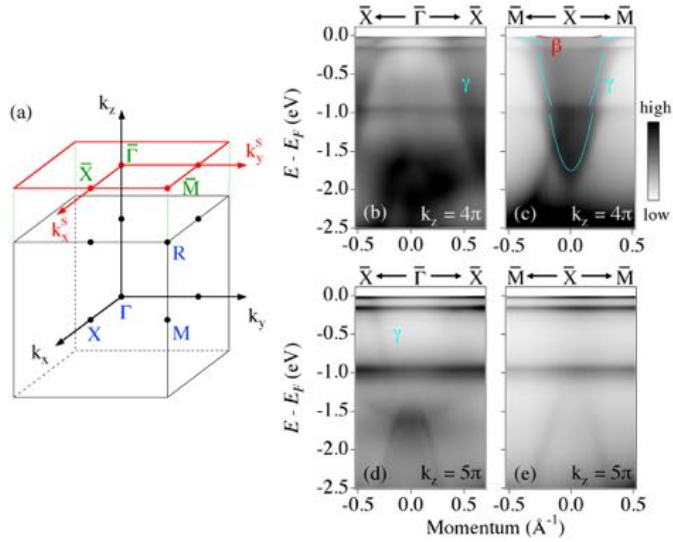

**Figure 5**. (Color online.) Valence band structure of $SmB_6$. (a) The first Brillouin zone of $SmB_6$ and the projection on the cleaving surface. High-symmetry points are also indicated. (b)-(c) ARPES intensity plots at $\bar{\Gamma}$ and $\bar{X}$ for the $k_z = 4\pi$ plane, respectively. (d)-(e) Same as (b)-(c), but for the $k_z = 5\pi$ plane. Reprinted with permission from Ref. [54].

VUV-ARPES is an ideal technique that can visualize the surface states of insulating $SmB_6$ to reveal their topological nature. **Figure 5** shows the ARPES spectra along high symmetry lines in SmB$_6$, reported by **Xu *et al.*** [54]. At low temperature ($T$ = 17 K), a highly dispersive electron-like band (γ band) hybridizes with three flat $4f^6$ bands, consistent with previous studies [55,56]. The γ band appears at $\bar{X}$ for k$_z$ = 4π and at $\bar{\Gamma}$ for k$_z$ = 5π, which are equivalent to the X point in the bulk Brillouin zone (BZ), indicating its strong bulklike 3D character. The hybridization between the γ band and *f* electrons opens a bulk gap Δ = 20 meV **(Fig. 5b)**, which agrees fully with the bulk band from theoretical calculations [39,40,57,58]. This leads to bulk insulating behavior in SmB$_6$ at low temperatures. On the other hand, 2D metallic states are observed in the hybridization band gap **(Fig. 6b)** and have three electron-like FSs in the first surface Brillouin zone (SBZ) **(Fig. 6a)**: one α pocket is centered at the $\bar{\Gamma}$ point and two β pockets encircle the $\bar{X}$ point. Similar ARPES results have reported by **Neupane *et al.*** [59] and **Jiang *et al.*** [60].

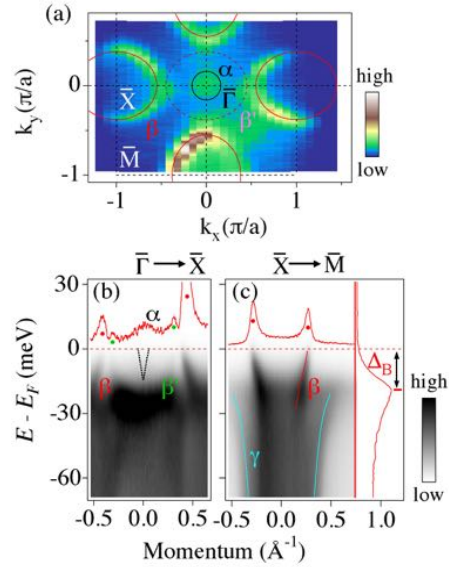

**Figure 6**. (Color online.) Plots of the Fermi surface (a), and ARPES intensity near the Fermi energy (E$_F$) along the $\bar{\Gamma}$-$\bar{X}$ (b) and $\bar{X}$-$\bar{M}$ (c) directions, measured at T= 17 K with hν = 26 eV and circular polarized light. Reprinted with permission from Ref. [54].

### 2.3 Two-dimensionality and surface origin of the in-gap states

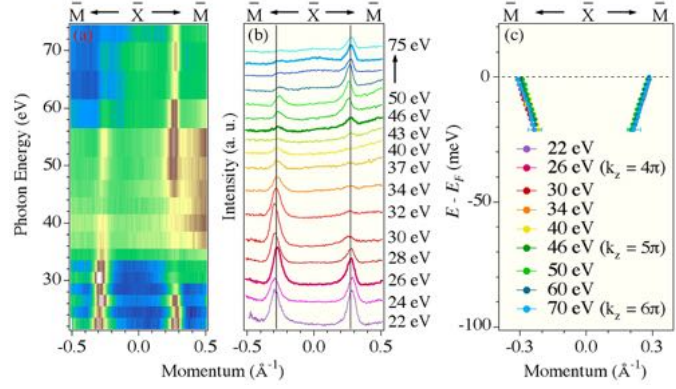

**Figure 7**. (Color online.) Photon energy dependent ARPES measurements. (a) Photon energy dependent intensity along the $\bar{X}$-$\bar{M}$ direction, covering more than 1.5 BZs along $k_z$. (b) Corresponding momentum distribution curves (MDCs) at $E_F$. (c) Extracted dispersions of the β band for different photon energies. Data are collected at T = 17 K with circular polarized light. Reprinted with permission from Ref. [54].

These in-gap states are distinct from the calculated bulk states[39,40,57,58]. The photon energy dependence of the ARPES spectrum further confirms its surface origin[54,59,60]. As seen from **Fig. 7a-b**, for the β band at the $\bar{X}$ point, the $k_F$ remains constant while the incident photon energy is varied. The extracted dispersions of the β band for various photon energies overlap each other (**Fig. 7c**), demonstrating the 2D nature of the β band. Similar measurements have been performed for the α band and, although the intensity of this this band is weak, the obtained results[54,60] reveal that the α band is also non-dispersive along $k_z$. The 2D nature of these in-gap states, which is not expected in the cubic symmetry of the bulk and totally different from the bulk γ band, indicates the surface origin of the α and β bands.

## 2.4 Surface reconstructions

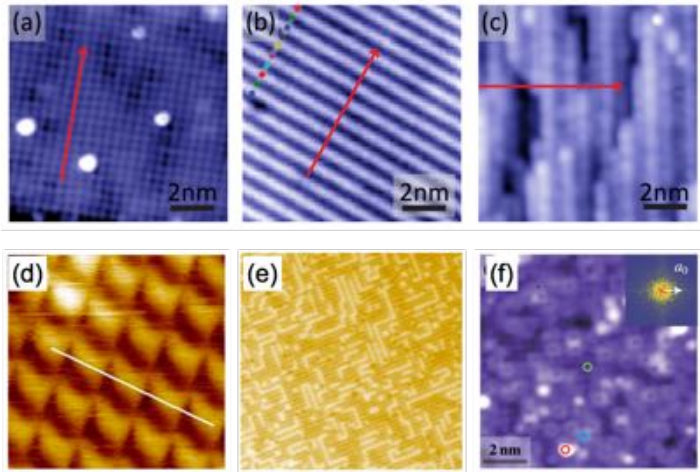

**Figure 8**. (Color online.) STM topographic images on the cleaved (001) surface of $SmB_6$. (a) 1×1 Sm termination (10 nm × 10 nm). (b) 1×2 half-Sm termination (10 nm × 10 nm). (c) Disordered B termination (10 nm × 10 nm). (d) 1×1 B termination (2 nm × 2 nm). (e) Complex reconstructed surface (30 nm × 30 nm). (f) Doughnut-like surface. (a)-(c) reprinted with permission from Ref. [61], (d)-(e) from Ref. [62], (f) from Ref. [63].

The surface of $SmB_6$ is complex because of the absence of a natural cleavage plane in the cubic crystal structure. Scanning tunneling microscopy (STM) studies by **Yee *et al*** [61] and **Rößler *et al*** [63] suggest that only a small amount of the cleaved (001) surface shows pristine Sm/B-terminations because these non-reconstructed surfaces are charge-polar and energetically unfavorable (**Fig. 8a, d**). Instead, a 1×2 reconstruction with half samarium termination, which is charge-neutral, appears over most of the surface (**Fig. 8b**). More complex surfaces are also observed by different cleaving methods, including the doughnut-like one of **Ruan *et al*** [62] (**Fig. 8f**).

STM studies[61,62,63] consistently show that the domain sizes at surfaces under any condition are much smaller than the footprint of the light in ARPES measurements (typically tens of micrometers). Therefore ARPES cannot resolve the electronic structures of a single surface domain; rather it picks up a summation of contributions from different surface domains. Nevertheless, the position dependence of the angle-integrated DOS from PES could be due to different surface conditions, as observed by **Denlinger *et al***[64].

Surface band structures are only well-defined for an ordered surface with translation symmetry. Therefore, the sharp and clear band structure observed in ARPES experiments is mainly from ordered surfaces with the signature of folding bands β' at the $\bar{\Gamma}$ point – i.e., the 1×2 reconstructed Sm-termination, as shown in **Fig. 6b** [54,60,65] This is consistent with low-energy electron diffraction (LEED) results with a space resolution in the same order, showing a superlattice structure of 1×2 reconstruction in addition to the 1×1 of the bulk[55].

## 2.5 Temperature evolution of electronic structure

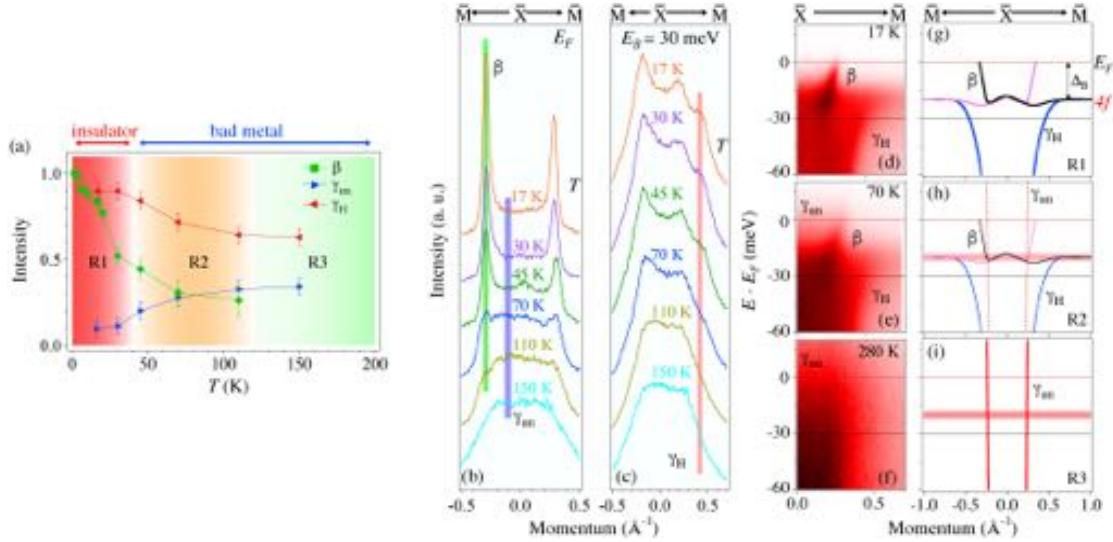

**Figure 9**. (Color online.) Temperature dependent ARPES measurements with circular polarized light. (a) Temperature dependence of the intensity of the surface state β band and the bulk conduction band with (without) hybridization with the *f* electrons $\gamma_H$ ($\gamma_{un}$). (b)-(c) MDCs along the $\bar{X}$-$\bar{M}$ direction as a function of temperature, taken at $E_F$ and $E_B = 30$ meV, respectively. (d)-(f) ARPES intensity plots for 17, 70, and 280 K. (g)-(i) Corresponding illustration of the band structures in the R1, R2, and R3 regions. Reprinted with permission from Ref. [65].

The evolution of both bulk and surface electronic structures as a function of temperature has been investigated by VUV-ARPES measurements[54,60]. Neither the surface α and β bands nor the bulk band gap resulting from the hybridization between the γ band and flat *f* band is observed at high temperatures, and they appear when T < 110 K. On the other hand, the gap appears to close at much lower temperatures (15~30 K), as shown in the DOS obtained from both partially angle-integrated laser-based ARPES[60] and STM[61] experiments. A quantitative temperature-dependent ARPES study by **Xu et al.** [65] reconciles the apparent controversy. A crossover from the high-temperature metallic phase to the low-temperature Kondo insulating phase takes place over a wide temperature region (R2 in **Fig. 9a**). In the crossover region a conduction band ($\gamma_{un}$ in **Fig. 9b**) crossing $E_F$ co-exists with a hybridized $\gamma_H$ band (**Fig. 9c**) with the ratio of $\gamma_{un}/\gamma_H$ reducing with decreasing temperature. The surface in-gap states emerge only when the *d-f* hybridization occurs (β in **Fig. 9a**). The partial $\gamma_{un}$ band crossing $E_F$ makes $SmB_6$ a bad metal in the R2 region, which results in the non-zero DOS at $E_F$ observed in angle-integrated ARPES[60] and STM[61]. Below 30 K, the *d-f* hybridization opens a clear gap, which turns the system into a bulk insulator. At very low temperatures (below 3.5 K), the surface in-gap states dominate the transport properties, causing the resistivity to saturate

54

instead of diverging as the temperature approaches zero.

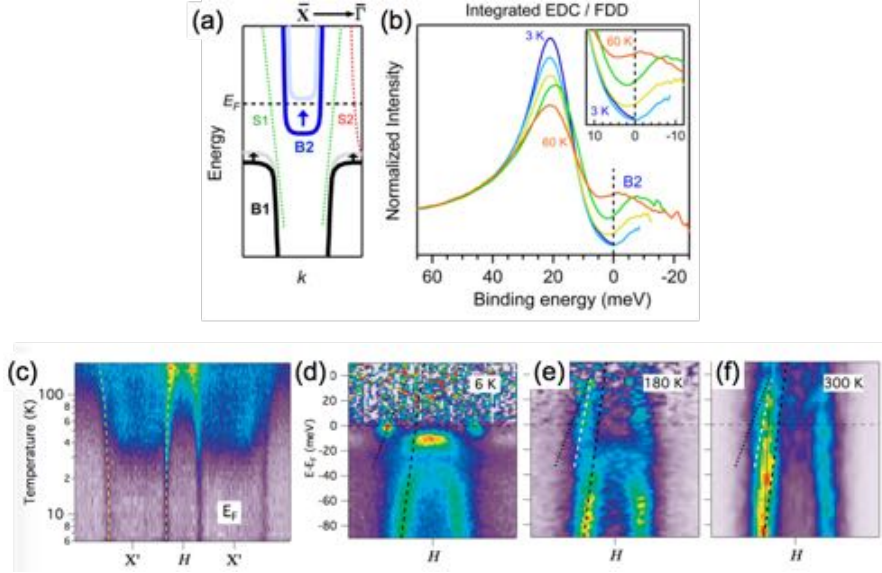

**Figure 10**. (Color online.) (a) Schematic diagram of hybridization band shifting in the crossover region. (b) Temperature dependent energy distribution curves divided by the Fermi-Dirac distribution, with enlarged spectra near $E_F$ shown in the inset. (c) Temperature-dependent ARPES intensity at $E_F$. (d)-(f), Band structure along X'-H-X' s at T = 6 K, 180 K and 300 K. (a)-(b) reprinted with permission from Ref. [66], (c)-(f) from Ref. [67].

The crossover scenario has also been proposed by **Min et al.**[66] and **Denlinger et al.**[67] Based on the observation that the energy position of the bottom (top) of the bulk hybridized conduction (valence) band shifts with changing temperature, Min *et al*. propose that charge-fluctuations in the bulk states occur in the crossover region (**Fig. 10a-b**). **Denlinger *et al*.** have observed a continuous change of the Fermi momentum (**Fig. 10c**) and Fermi velocity (**Fig. 10d-f**) when temperature varies in the cross-over region, and at the same time the in-gap states evolve from 2D states at low temperature to 3D at high temperature.

## 3. Scenarios for the in-gap states in $SmB_6$

### 3.1 Topological Kondo insulator surface states

#### 3.1.1 Consistency with TSS calculations

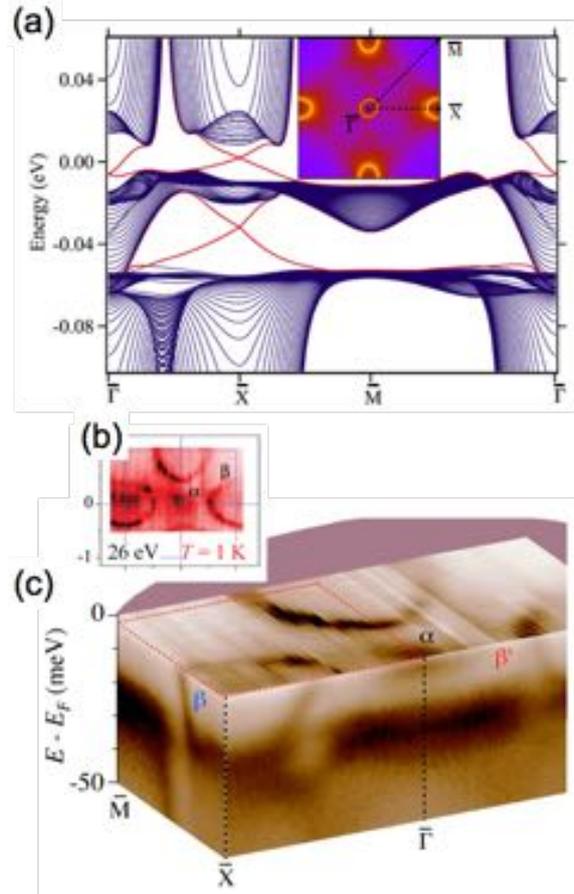

**Figure 11**. (Color online.) Comparison between theoretical calculations and experimental results of the band structures on the $SmB_6$ (100) surface. (a) Theoretical results of band structure obtained by LDA+Gutzwiller method. The red lines represent the TSS. The inset shows the Fermi surfaces for the TSS. (c) Experimental Fermi surface of the surface in-gap states. (d) 3D plot of the electronic structure of $SmB_6$ measured by ARPES. (a) reprinted with permission from Ref. [57], (b) from Ref. [65], (c) from Ref. [68].

The experimentally determined surface states in refs. [54,59,60] are consistent with the topologically non-trivial surface states predicted by the calculations, in which strong correlation effects are taken into account. **Figure 10a** shows the dispersions of the topological surface states (red lines) and the projections of the bulk bands on the (100) surface (blue lines) obtained from the local-density approximation combined with the Gutzwiller method by **Lu et al.**[57]. The odd number of the Fermi surfaces of the surface states predicated by the calculation, as well as their shapes and locations, are consistent with the ARPES measurements (**Fig. 11b-c**), which provide the first spectroscopic evidence that at low temperatures the surface states in the TKI candidate $SmB_6$ are topologically non-trivial in nature. Similar calculation results have also been obtained by Takimoto[58] and Alexandrov et

56

al.[69]. In spite of these agreements between the experiments and calculations, the Fermi velocity of the β band obtained by ARPES measurements[54,59,60] is one order of magnitude larger than that of the calculations[39,40,57,58], which may result from the Dirac points buried in the intense bulk *f*-/*d*-bands, making the Fermi velocity difficult to extract from the experiments. The higher Fermi velocity found in ARPES measurements can be reconciled by taking the possible Kondo break-down effect on the surface into account, as shown in the recent calculations by Alexandrov et al.[70] **(Fig. 12b).** The Fermi surface areas calculated by Alexandrov et al. are also larger than that obtained from the uniform solution (**Fig. 12a**), which reproduces well the surface in-gap state dispersions obtained from ARPES.

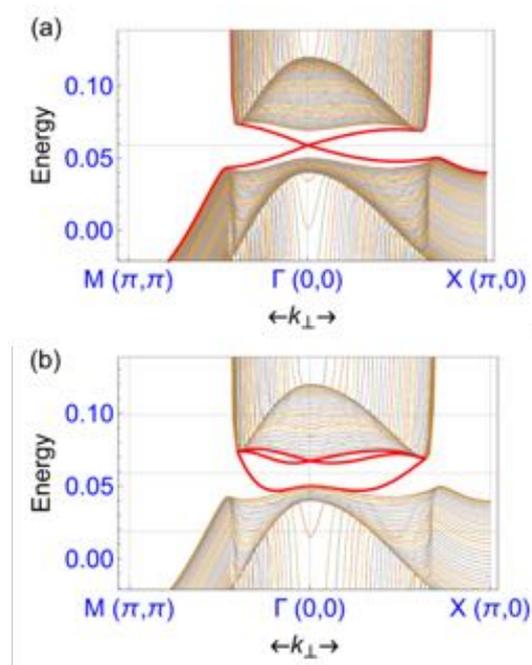

**Figure 12**. (Color online.) (a) Band structure calculation for the uniform solution. (b) Band structure calculation for the surface Kondo breakdown with a much larger Fermi surface and higher velocity quasiparticles, which are consistent with experimental results. Reprinted with permission from Ref. [70].

The surface in-gap states in $SmB_6$ show unusual properties, which make it distinct from other trivial surface states and support the TKI scenario.

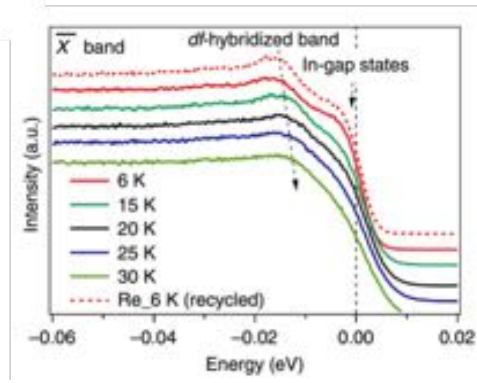

**Figure 13**. (Color online.) Temperature-dependent density of states integrated near the $\bar{X}$ point. Reprinted with permission from Ref. [59].

### 3.1.2 Robustness against thermo-cycling.

Most trivial surface states are sensitive to the surface conditions and unlikely to survive thermo-cycling. In contrast, because they are protected by time-reversal symmetry and encoded by the bulk wave functions, topologically non-trivial surface states are expected to be robust against the thermo-cycling process. In measurements on $SmB_6$, **Neupane** et al. have shown that the ARPES spectral intensity of the in-gap surface state shows no obvious change after thermo-cyclings (**Fig. 13**)[59], which provides further evidence that these surface states are topologically non-trivial in nature.

### 3.1.3 Effects of magnetic/non-magnetic impurities.

**Kim** *et al*. studied the effects of impurities in thickness-dependent transport measurements[71]. The results show that resistivity saturation behavior at low temperature, which is directly related to the surface in-gap states, is robust against non-magnetic Y impurities (**Fig. 14a**) but destroyed by a similar amount of magnetic Gd impurities (**Fig. 14b**). The sensitivity of conductance at low temperature to the perturbations that break TRS also supports the scenario that the in-gap surface states originate from the $Z_2$ topological feature of the bulk states in $SmB_6$.

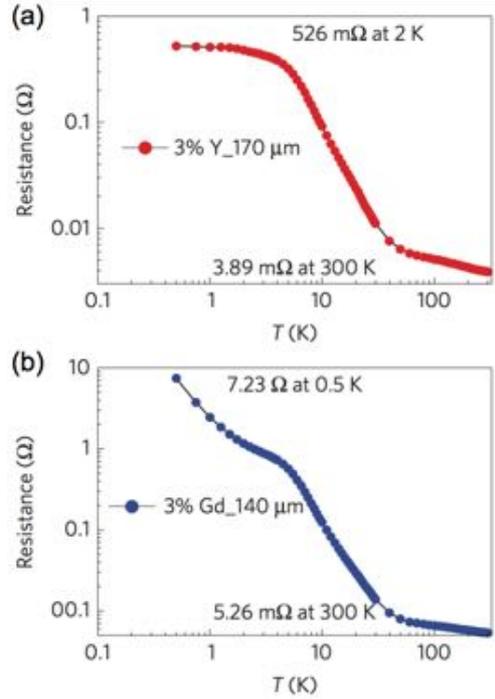

**Figure 14**. (Color online.) Temperature dependent resistance curves of Y- and Gd-doped $SmB_6$. Reprinted with permission from Ref. [71].

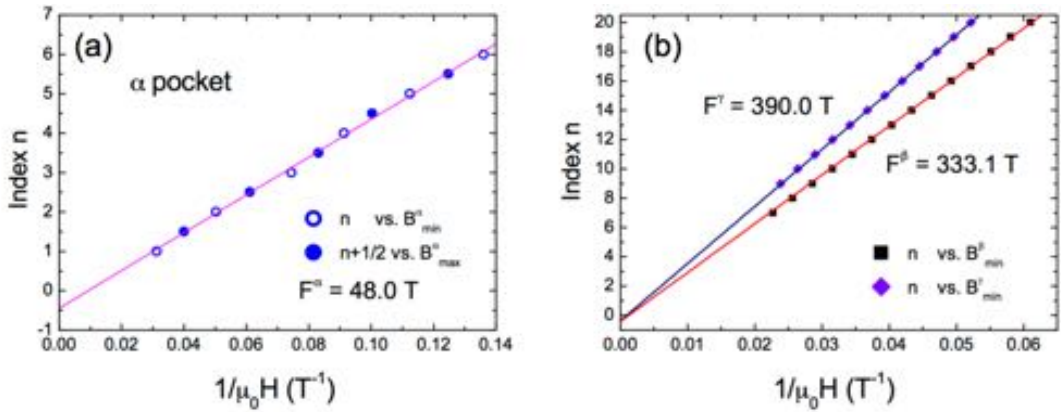

**Figure 15**. (Color online.) Plot of the Landau Level index of the surface in-gap states vs. magnetic field for α (a), β and γ (b). Reprinted with permission from Ref. [72].

### 3.1.4 Evidence for Dirac cone dispersion of the surface states from quantum oscillations.

The first magneto-torque oscillation results from the (100) and (110) surface states in $SmB_6$ were reported by **Li** et al.[72]. The obtained angular dependence of the oscillating frequency F is fitted to $1/\cos(\theta-45°)$, where $\theta$ is the angle between magnetic field and crystalline c axis, suggesting the quantum oscillation signals are related to 2D electronic states. The -1/2 Landau

Level index obtained from the extrapolation to the high magnetic field limit indicates that the in-gap states have 2D Dirac cone-like dispersion (**Fig. 15**). On the other hand, **Tan** et al. have also observed magneto-torque oscillations in $SmB_6$ but attributed the quantum oscillation signals to 3D bulk states[73]. However, it is a mystery how these quantum oscillation signals can occur in a bulk insulating state. **Denlinger** et al. analyzed the quantum oscillation data from the both groups, and favored the 2D surface states by comparing with their ARPES results on (100) and (110) surface[74]. **Erten** et al. pointed out that the Kondo breakdown at the surface has to be considered in order to interpret the quantum oscillation results, which can reconcile the contradiction in the topological surface states scenario[75].

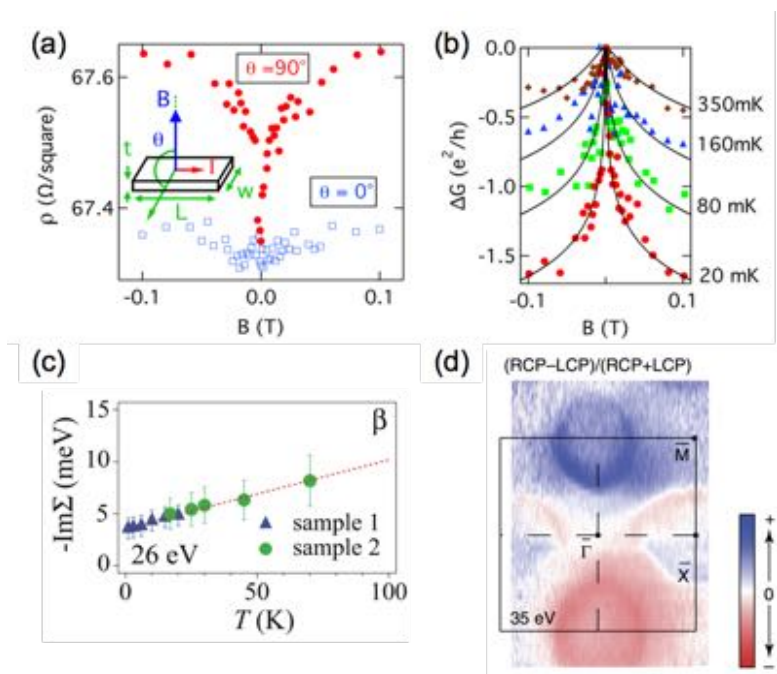

Figure 16. (Color online.) (a) Magneto-resistance with magnetic fields (both perpendicular and in-plane direction) measured at 20 mK. (b) $\Delta G = \Delta(1/\varrho)$ fitted to Hikami-Larkin-Nagaoka equation. (c) Temperature dependence of imagery part of self-energy by analysis of the MDC fits. (d) Circular dichroism measurements of the Fermi surface. (a)-(b) reprinted with permission from Ref. [77], (c) form Ref. [65] and (d) from [60].

### 3.1.5 Evidence for spin-polarized surface in-gap states.

The most fundamental difference between a TSS and a trivial surfaces state is that a TSS forms an odd number of Fermi surfaces that are spin polarized with spin-momentum locking. Therefore, the spin-polarization of the surface in-gap states is compelling evidence for the TKI scenario in SmB$_6$. This is supported by observation of weak localization (WAL)[76] in SmB$_6$ by **Thomas** et al.[77]. For a TSS with spin-momentum locking, the destructive interference effect between electrons with time-reversed paths will lower the resistance [37,38,78,79], i.e., WAL. Due to TRS breaking, this effect is destroyed in the presence of a magnetic field, leading to a resistance peak at zero magnetic field, which is observed in SmB$_6$ and well fitted to the Hikami-Larkin-Nagaoka (HLN) equation (**Fig. 16a-b**). It is worth mentioning that strong spin-orbit coupling can also induce WAL, which implies that the observed WAL in SmB$_6$ cannot yet be conclusively linked to a TSS. The spin-polarized surface states are also suggested by the linear, non-Fermi liquid temperature dependence of the quasiparticle scattering rate[65], as shown in **Fig. 16c**. Similar scattering rate suppression due to spin-momentum locking is also observed in non-interacting TI Bi$_2$Se$_3$[80]. Circular dichroism results by **Jiang** et al.[60] suggest that the in-gap states possess chirality of the orbital angular momentum (**Fig. 16d**), providing further indirect evidence for the spin-polarized TSS. In the last part of this review, we discuss the direct observation of the spin texture of the surface in-gap states in SmB$_6$ as strong evidence of the TKI scenario[68].

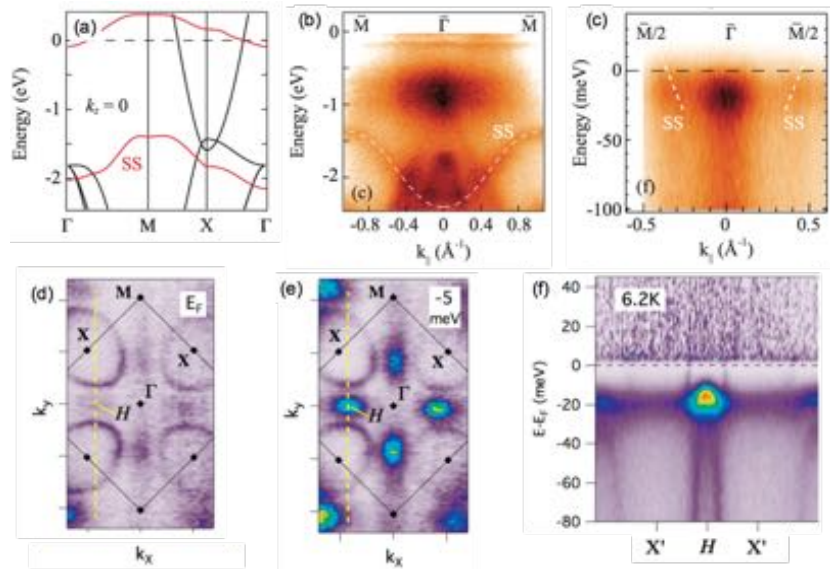

**Figure 17**. (Color online.) (a) Bulk band structure calculation (black lines) with polar surface states (red lines). (b) ARPES results along $\bar{M}$-$\bar{\Gamma}$-$\bar{M}$ direction. (c) Zoomed in near E$_F$. (d)-(e) Constant energy maps at E$_F$ and -5 meV. (f) ARPES results along the $\bar{X}$-$\bar{H}$-$\bar{X}$ direction. (a)-(c) reprinted with permission from Ref. [81], (c)-(f) form Ref. [67].

### 3.2 Trivial surface states

Trivial polarity-driven surface states have been proposed by **Zhu** et al. based on their ARPES and density functional theory studies[81]. Unlike other theoretical approaches[39,40,57,58], the Sm 4f states were treated as core levels that do not interact/hybridize with the conduction/valence electrons. The calculation reveals a metallic bulk band and two trivial surface state bands on the non-reconstructed (100) surface of SmB$_6$ (**Fig. 17a**); one located at 1.5 eV below E$_F$ and the other located near the chemical potential, forming a large electron-like pocket at the SBZ center $\bar{\Gamma}$ point. The calculated trivial surface states are supported by their ARPES spectra along the $\bar{M}$-$\bar{\Gamma}$-$\bar{M}$ direction (**Fig. 17b**). However, this trivial polarity-driven surface state with a single electron-like pocket, as seen from **Fig. 17c,** is not observed in other ARPES studies [54,59,64,65,66,67], one possible reason might be the rarity of the non-reconstructed (100) surface and its small domain sizes, as suggested by STM results [61,62,63]. On the other hand, **Denlinger** et al.[67] observed that spectral intensity near the chemical potential forms four individual hot spots (**Fig. 17d-e**), and the momentum locations of the spots along the $\bar{\Gamma}$-$\bar{M}$ direction are similar to those of the trivial surface state suggested by **Zhu** et al. **Denlinger** et al. proposed a different scenario that these hot spots are from intensity tails of bulk 4f states gapped at low temperature (**Fig. 17f**), instead of trivial polar metallic surface states. There is also a scenario that the metallic states around $\bar{X}$ could result from the chemical potential at the surface and near=surface regions being different inform the bulk. In this scenario, they are not in-gap states[82].

### 3.3 Rashba surface states

**Hlawenka** *et al*. reported ARPES results on the B$_6$-terminated (100) surface in SmB$_6$[83]. In addition to the two β pockets around the $\bar{X}$ point observed by other groups' ARPES results on the (100) surface, they observed three electron-like pockets centered at the $\bar{\Gamma}$ point on the B-termination surface. They have attributed the inner two electron-like pockets around the $\bar{\Gamma}$ point to Rashba-split surface states, and the outer one to the folding of band β on the 1×2 reconstructed surface. On the other hand, this surface assessment is inconsistent with observations in STM experiments that do not show any 1×2 reconstruction on the B$_6$-termination in SmB$_6$[61,62,63]. The B$_6$-terminated surface revealed by STM is not reconstructed and has a domain size of around 10 nm, which is much smaller than the spatial resolution of ARPES. Indeed, the dispersion of the outer pocket at the $\bar{\mathit{\Gamma}}$ point shows great similarity to the folded band β' on the 1×2 reconstructed Sm termination[54,60]. Currently, the observation of Rashba surface states is still under debate, and further studies are needed.

## 4. Spin texture of SmB$_6$

### 4.1 Spin-resolved ARPES

For systems in which the spin degeneracy is lifted by the TRS breaking (magnetic materials) or inversion symmetry breaking (non-centrosymmetric materials, surfaces and interfaces), the spin polarization of the electronic states is critical to understand the underlying physics. ARPES with spin resolution (SARPES) is one of the most powerful techniques that can directly measure the spin polarization of the wave functions of a crystal. In a SARPES measurement, photoelectrons with well-defined kinetic energy and emission angle (corresponding to the desired initial states $E_B(k)$) are collected like in ARPES, and then the collected electron beam is sent to a spin detector which measures the difference of currents for the spin-up and spin-down. Most spin detectors, including spin-polarized low energy electron diffraction (SPLEED)[84], very low energy electron diffraction (VLEED) and Mott detector[85], are based on the asymmetry of the spin-polarized electron scattering in crystalline solids. Compared to the former two spin detectors, Mott detectors have the advantages of high stability and high accuracy, and are the most commonly used for SARPES. As originally discussed by Mott, when electrons with high kinetic energy (≥25 keV) scatter off heavy nuclei (e.g., gold foil), the scattering rate shows a dramatic dependence on the spin polarization direction. The asymmetry is written as: $A = (N_+ - N_-)/(N_+ + N_-)$, where $N_+$ and $N_-$ are the numbers of electrons parallel and anti-parallel to the spin polarization direction, respectively. The spin polarization rate (P) is proportional to the asymmetry: $P = A/S$, where a coefficient called the Sherman function (S)[86] can be determined by measuring a fully polarized electron beam. With the measured spin polarization rate, the spin-resolved ARPES spectra can be expressed as: $I_\uparrow = (1 + P)(N_+ + N_-)/2$; $I_\downarrow = (1 - P)(N_+ + N_-)/2$, where $I_\uparrow/I_\downarrow$ is the SARPES intensity for spin up/down. For systems with strong spin-orbit coupling, the states can possess spin structures in which a single quantization axis of the spins cannot be defined so that measurements of the spin components along various axes are required to determine the polarization directions of the photoelectrons. In practice, two pairs of electron detectors are installed orthogonally within a single Mott detector in order to measure the spin polarization along two different directions simultaneously.

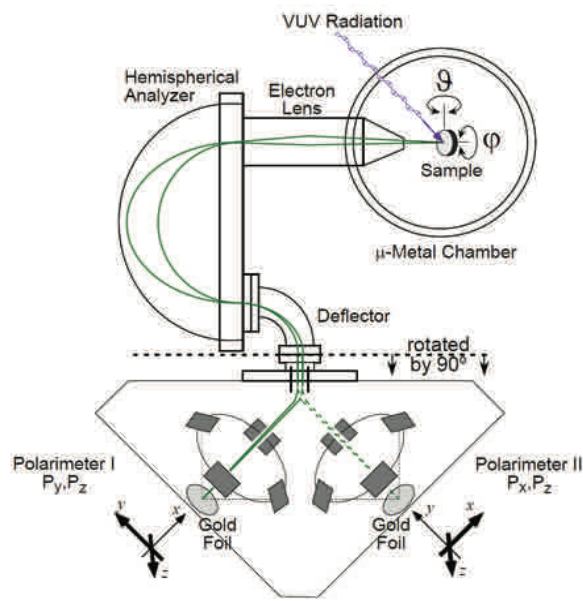

**Figure 18**. (Color online.) Schematic diagram of COPHEE, the COmplete PHotoEmission Experiment. Reprinted with permission from Ref. [87].

**Figure 18** shows the schematic of the COmplete PHotoEmission Experiment (COPHEE) setup at the Surface/Interface Spectroscopy beamline at the Swiss Light Source[87]. The photoelectrons, selected by the hemispherical analyzer with desired energy and emission angle, are accelerated to 40 keV, bent by $\pm 45°$, and sent into two orthogonal Mott detectors. The geometry of the experiment involving two Mott detectors allows the determination of the complete spin polarization of the photoelectrons along all three special axes. The spin-polarization can be further transformed into the sample frame $P_x$, $P_y$, and $P_z$, by a simple rotation matrix. Corresponding spin-resolved ARPES spectra can be calculated.

### 4.2 Spin texture of the surface in-gap states in SmB$_6$

The determination of the spin texture of the surface in-gap states by SARPES makes it possible to distinguish the TSS from the trivial surface states and provides key evidence for examining the TKI scenario in $SmB_6$. However, it is extreme challenging, because (1) the efficiency of SARPES is normally five orders of magnitude lower than that of ARPES due to the scattering process and single-detection-channel in the Mott detector, so that much more time is needed to obtain reliable data; (2) the energy resolution of synchrotron-based SARPES at the present stage is 60~150 meV, while the surface in-gap states in $SmB_6$ are within the bulk Kondo gap of ~ 20 meV; (3) the bulk $f$ states and conduction d states are intense and located close to the surface in-gap states.

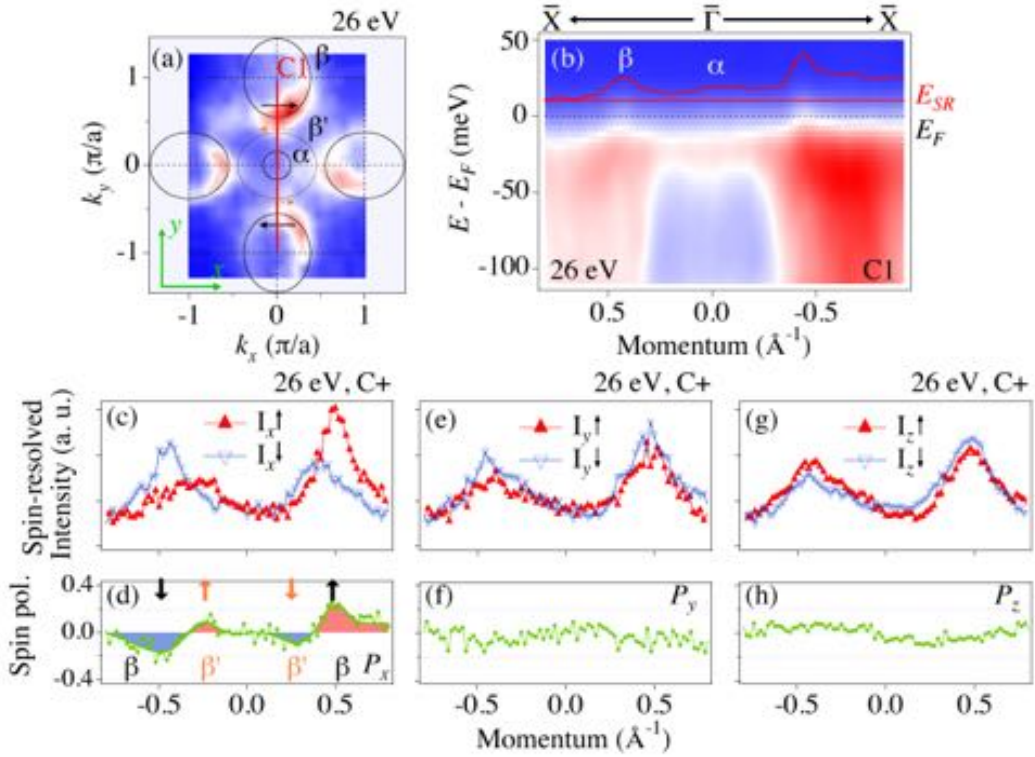

**Figure 19**. (Color online.) (a) FS map of $SmB_6$. (b) Low energy excitations near $E_F$ along C1 in a. The red curve is the MDC at 10 meV above E($E_{SR}$). (c) Spin-resolved MDCs along the x direction measured at $E_{SR}$, with a photon energy of 26 eV and C+ polarization. (d) Corresponding spin polarization along the x direction. (e)-(f) Same as c,d but along the y direction. (g,h) Same as c,d but along the z (out-of-plane) direction. Reprinted with permission from Ref. [68].

By pushing the resolution of SARPES to the limit (60 meV), **Xu** et al. succeeded to measure the spin texture of the Fermi surface of the surface in-gap states in $SmB_6$. The spin-integrated FS mapping (**Fig. 19a**) and the band structure along the $\bar{M}$-$\bar{\Gamma}$-$\bar{M}$ direction (**Fig. 19b**) clearly show the surface bands α and β. To determine the contribution from the bulk states, spin resolved MDC measurements are taken at $E_{SR}$ (10 meV above $E_F$). As shown in **Fig. 19**c,d, The difference between $I_\uparrow$ and $I_\downarrow$ of two branches of the β band clearly reveal that the surface bands are spin-polarized along the x direction. Two additional peaks are observed between the main β band peaks in the $P_x$ spectrum (**Fig. 19d**) and are attributed to the folded band β' originating from the 1×2 surface reconstruction. On the other hand, the β and β' bands show no spin polarization along the y and z direction as seen in **Fig. 19e-h**.

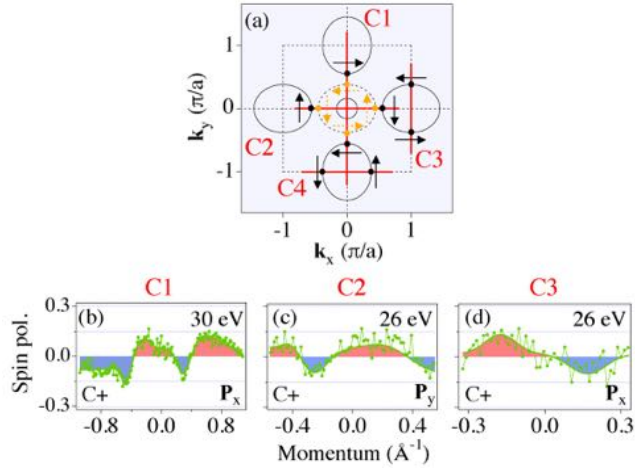

**Figure 20**. (Color online.) (a) Schematic diagram of the spin polarized Fermi surface of $SmB_6$. (b) Spin polarization (x component), measured along C1 with hν = 30 eV. (c) Spin polarization (y component), measured along C2 with hν = 26 eV. (d) Same as (c), but for the x component measured along C3. Reprinted with permission from Ref. [68].

More SARPES measurements have been done to explore the complete spin texture of the surface in-gap states (**Fig. 20**b-d). As summarized in **Fig. 20**a, the measured spin polarization of the β band shows a helical structure with spin-momentum locking, fully consistent with both TRS and the crystal symmetry. The folded band β' has the same spin texture as the original β band, which is expected from an Umklapp mechanism[88].

### 4.3 Photon energy- and polarization-dependence of the spin polarization of the surface in-gap states

66

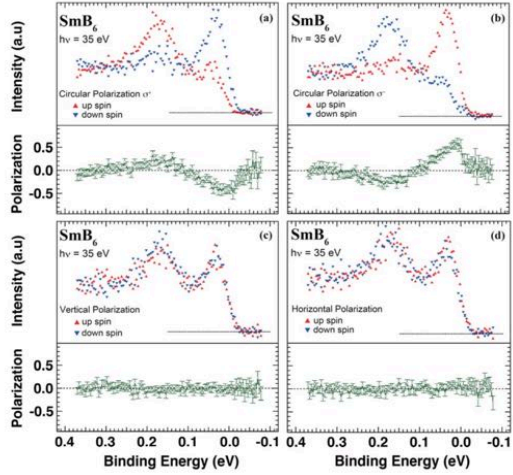

**Figure 21**. (Color online.) (a)-(d)Spin-polarized energy distribution curves measured with different polarizations (C+, C-, linear horizontal and linear vertical). Reprinted with permission from Ref. [93].

In SARPES on states with no net spin polarization, spin polarization signals may appear due to photoemission effects[89,90]. These effects have been observed in the core-levels of non-magnetic solids[91], the bulk valence bands of TIs[92] and the bulk f states of $SmB_6$[93]. Unlike the intrinsic spin signal, the non-intrinsic spin signal strongly depends on the energy and polarization of incident photons. As seen in **Fig. 21**, the non-intrinsic signal of the f states in $SmB_6$ changes direction when the photon polarization changes from C+ to C- circularly polarized light and vanishes with linear polarization (l-pol). To confirm that the measured spin texture of the surface in-gap states is intrinsic, photon-energy and -polarization dependent SARPES measurements were carried out by **Xu el al**. For various incident photon energies (**Fig. 19d and Fig. 20a**), a consistent momentum-locked spin texture is obtained. The spin polarization of the photoelectrons also does not depend on photon polarization (**Fig. 22b-e**), confirming that the measured spin polarizations reflect the intrinsic spin structure.

### 4.4 Spin polarization of the bulk states

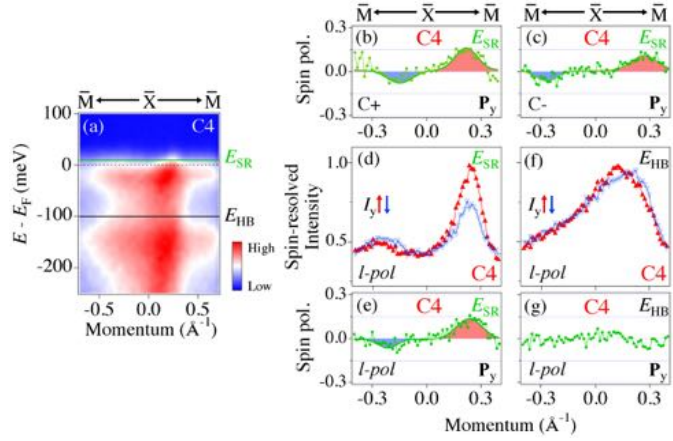

**Figure 22**. (a) Low-energy excitations near $E_F$ along the high symmetry line $\bar{M}$-$\bar{X}$-$\bar{M}$. (b) Spin polarization (x component), measured along C1 with hv = 30 eV. (c) Spin polarization (y component), measured along C2 with hv = 26 eV. (d) Same as (c), but for the x component measured along C3. Reprinted with permission from Ref. [68].

In order to determine the contribution of the bulk *f* and *d* states to the spin-polarization signal, SARPES measurements have been done. The bulk *f* states show no spin signal with l-pol light (**Fig. 21c-d**) as mentioned previously. Measurements at a higher binding energy ($E_{HB}$), where the bulk d states dominate the photoemission intensity, also show no spin polarization (**Fig. 22f-g**). Therefore, the spin texture observed in the SARPES measurements originates from the surface in-gap states, which provides compelling evidence that $SmB_6$ is the first realization of a TKI.

### 4.5 Spin texture calculations on $SmB_6$

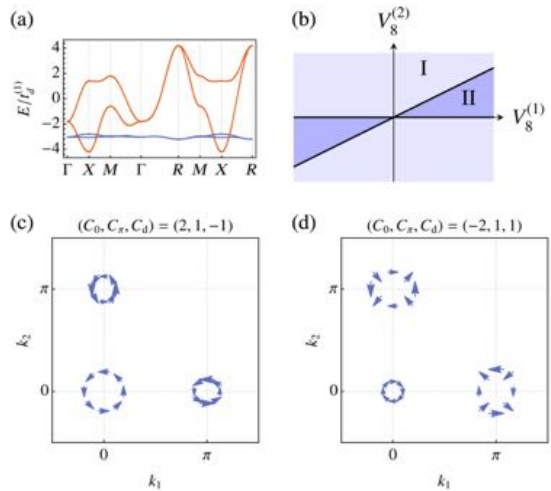

**Figure 23**. (Color online.) (a) Bulk band structure calculation without hybridization. (b) The phase diagram. (c)-(d) The two spin textures in phases I and II, respectively. Reprinted with permission from Ref. [97].

The spin texture of the TSS in SmB$_6$ **(Fig. 23d)** proposed in theoretical studies[94,95,96] conflicts with that obtained in SARPES measurements[68], which has the opposite winding numbers at the $\bar{X}$ point. **Legner** et al. pointed out that the relative strength of the hybridization parameters has to be taken into account in the effective model[97]. They obtained two types of spin textures of the TSS distinguished by the surface mirror Chen numbers; one **(Fig. 23d)** is the same as in other calculations[94,95,96] and the other **(Fig. 23c)** is fully consistent with the experimental spin texture[68]. A phase diagram **(Fig. 23b)** is obtained by calculating the winding numbers at the $\bar{X}$ point in their model. Under the same $Z_2$ topological invariants, the system can undergo topological phase transitions with variations of the TSS spin texture. Their work not only reconciles the contradiction between the experimental and theoretical spin texture of the TSS in SmB$_6$, serving as strong evidence for the TKI scenario, but also can be generally applied to other TI systems. Similar result was also reported by **Baruselli** et al.[98]

## 5. Summary and outlook

ARPES provides direct visualizations of the metallic surface states in insulating SmB$_6$ at the low temperature. The surface metallic states show very distinct behavior from trivial surface states in terms of the robustness, temperature dependence, chirality of the orbital angular momentum, and especially the helical spin-texture (β band). The great consistency in transport, other spectroscopy and theoretical studies of the TSS strongly support the realization of a TKI in SmB$_6$. The strong correlation effects and diverse surface conditions make the system extremely complicated. There are still open questions such as the effects of the sample growth conditions (small amount of impurities and disorder) on the resistivity behavior at low temperature. The dispersion of the α band and its termination dependence is not clear. The spin texture of the α band and the STM quasiparticle interference patterns, which can conclusively pin down the TKI scenario, are still missing. **Syers** et al. demonstrate control of bulk and surface conduction by gating thin films, representing a step forward in applications of the robust metallic states in SmB$_6$[99]. Finding other TKI candidates, especially with a high Kondo temperature, will advance the current understanding of correlated TI's and applications of TSS's in the future.

**Acknowledgement**

We acknowledge valuable discussions with Zhong Fang, Kai Sun, Hugo Dil, Manfred Sigrist and Joel Mesot. We also thank Jihwey Park, Nicholas C. Plumb and Binbin Fu for the help in preparing the manuscript. This work was supported by NCCR-MARVEL funded by the Swiss National Science Foundation and the Sino-Swiss Science and Technology Cooperation (No. IZLCZ2138954).

**References**

[1] X.-G. Wen, Int. J. Mod. Phys. B **4**, 239 (1990)
[2] X.-G. Wen, J. Mod. Phys. B **6**, 1711 (1992).
[3] K. V. Klitzing, G. Dorda and M. Pepper, Phys. Rev. Lett. **45**, 494 (1980).
[4] D. J. Thouless, M. Kohmoto, M. P. Nightingale, and M. den Nijs, Phys. Rev. Lett. **49**, 405 (1982).
[5] M. Kohmoto, Ann. Phys. (N.Y.) **160**, 343 (1985).
[6] J. E. Moore & L. Balents, Phys. Rev. B 75, 121306(R) (2007).
[7] V. L. Ginzburg and Ll D. Landau, Zh. Eksp. Teor. Fiz. **20**, 1064 (1950).
[8] C. L. Kane and E. J. Mele, Phys. Rev. Lett. **95**, 146802 (2005).
[9] C. L. Kane and E. J. Mele, Phys. Rev. Lett. **95**, 226801 (2005)
[10] B. A. Bernevig, T. L. Hughes and S. C. Zhang Science **314**, 1757 (2006)
[11] M. König, S. Wiedmann, C. Brüne, A. Roth, H. Buhmann *et al.*, Science **318**, 766 (2007).
[12] L. Fu & C. L. Kane, Phys. Rev. B **76**, 045302 (2007).
[13] Y. Xia, D. Qian, D. Hsieh *et al.*, Nat. Phys. **5**, 398 (2009).
[14] S. Souma, K. Kosaka, T. Sato *et al.*, Phys. Rev. Lett. **106**, 216803 (2011).
[15] C. L. Kane and E. J. Mele, Science **314**, 1692 (2006).
[16] Y. Ando, J. Phys. Soc. Jpn **82**, 102001 (2013)
[17] D. Hsieh, D. Qian, L. Wray *et al.*, Nature **452**, 970 (2008).
[18] H. -J. Zhang, C. –X. Liu, X. –L. Qi *et al.*, Nature Phys. **5**, 438 (2009).
[19] Y. L. Chen, J. G. Analytis, J.-H. Chu *et al.*, Science **325**, 178 (2009).
[20] Y. Xia et al., Nature Phys. **5**, 398 (2009).
[21] B. Yan, C.-X. Liu, H.-J. Zhang *et al.*, Europhys. Lett. **90**, 37 002 (2010).
[22] H. Lin, R. S. Markiewicz, L. A. Wray *et al.*, Phys. Rev. Lett. **105**, 036404 (2010).
[23] B. Rasche, A. Isaeva, M. Ruck et al., Nature Materials **12**, 422 (2013).
[24] W. Zhang, R. Yu, W. Feng *et al.*, Phys. Rev. Lett. **106**, 156808 (2011).
[25] H. M. Weng, X. Dai and Z. Fang, Phys. Rev. X **4**, 011002 (2014)
[26] X. L. Sheng, Z. J. Wang, R. Yu *et al.*, Phys. Rev. B 90, 245308 (2014)
[27] B. Yan, M. Jansen, C. Felser, Nature Phys. **9**, 709 (2013).
[28] H. -S. Kim, C. H. Kim, H. Jeong, H. Jin, and J. Yu, Phys. Rev. B **87**, 165117 (2013).
[29] T. Sato, K. Segawa, H. Guo *et al.*, Phys. Rev. Lett. **105**, 136802 (2010).

[30] K. Kuroda, M. Ye, A. Kimura, S. V. Eremeev *et al*., Phys. Rev. Lett. **105**, 146801 (2010).

[31] Y. Ando and L. Fu, Annu. Rev. Condens. Matter Phys **6**, 361(2015)

[32] L. Fu, Phys. Rev. Lett. **106**, 106802 (2011).

[33] Y. Tanaka, Zhi Ren, T. Sato *et al*., Nature Phys. **8**, 800 (2012).

[34] P. Dziawa, B. J. Kowalski, K. Dybko *et al*., Nature Materials **11** 1023 (2012).

[35] S.-Y. Xu, C. Liu, N. Alidoust *et al*., Nature Communications **3**, 1192 (2012).

[36] Y. Tanaka, T. Sato, K. Nakayama *et al*., Phys. Rev. B **87**, 155105 (2013).

[37] M. Z. Hasan and C. L. Kane, Rev. Mod. Phys. **82**, 3045 (2010).

[38] X. L. Qi and S. C. Zhang, Rev. Mod. Phys. **83**, 1057 (2011).

[39] M. Dzero, K. Sun, V. Galitski and P. Coleman, Phys. Rev. Lett. **104**, 106408 (2010).

[40] M. Dzero, K. Sun, P. Coleman and V. Galitski, Phys. Rev. B **85**, 045130 (2012).

[41] Maxim Dzero, Jing Xia, Victor Galitsk and Piers Coleman, Annual Review of Condensed Matter Physics, Volume 7 (2016).

[42] J. W. Allen, B. Batlogg, P. Wachter, Phys. Rev. B **20**, 4807 (1979).

[43] J. C. Cooley, M. C. Aronson, Z. Fisk, P. C. Canfield, Phys. Rev. Lett. **74**, 1629 (1995).

[44] N. E. Sluchanko, V. V. Glushkov, B. P. Gorshunov *et al.,* Phys. Rev. B **61**, 9906 (2000).

[45] D. J. Kim, S. Thomas, T. Grant, J. Botimer et al., Scientific Reports 3:3150 (2014)

[46] S. Wolgast, C. Kurdak, K. Sun et al., Phys. Rev. B **88**, 180405 (2013)

[47] X. H. Zhang, N. P. Butch, P. Syers et al., Phys. Rev. X **3**, 011011 (2013)

[48] H. Hertz, Ann. Phys. 267 (8) 983–1000 (1887).

[49] S. Hüfner, Photoelectron Spectroscopy: Principles and Applications, Springer, (2003).

[50] A. Damascelli, Phys. Scr. 2004 (T109) 61 (2004).

[51] P. Richard, T. Sato, K. Nakayama, T. Takahashi, H. Ding, Rep. Prog. Phys. 74 (12), 124512 (2011).

[52] M. P. Seah and W. A. Dench, Surface and Interface Analysis, 1, 2 (1979).

[53] V. N., Strocov, J. Electron Spectrosc. Relat. Phenom. 130, 65–78 (2003).

[54] N. Xu, X. Shi, P. K. Biswas, C. E. Matt, R. S. Dhaka, Y. Huang, N. C. Plumb, M. Radovic, J. H. Dil, E. Pomjakushina, K. Conder, A. Amato, Z. Salman, D. McK. Paul, J. Mesot, H. Ding, and M. Shi, Phys. Rev. B 88, 121102(R) (2013).

[55] H. Miyazaki, T. Hajiri, T. Ito, S. Kunii, and S. I. Kimura, Phys. Rev. B 86, 075105 (2012).

[56] S. K. Mo, G. H. Gweon, J. D. Denlinger, H. D. Kim, J. W. Allen, C. G. Olson, H. Hochst, J. L. Sarrao, and Z. Fisk, Phys. B 281–282, 716 (2000).

[57] F. Lu, J. Z. Zhao, H. Weng, Z. Fang and X. Dai, Phys. Rev. Lett. **110**, 096401 (2013).

[58] T. Takimoto, J. Phys. Soc. Jpn. 80, 123710 (2011).

[59] M. Neupane, et al., Nat. Commun. 4, 2991 (2013).

[60] J. Jiang et al., Nat. Commun. 4, 3010 (2013).

[61] M.M. Yee et al., arXiv: 1308.1085 (2013)

[62] W. Ruan et al., Phys. Rev. Lett. 112, 136401 (2014)

[63] S. Rößler et al., PNAS 111, 4798 (2014)

[64] J. D. Denlinger et al., arXiv: 1312.6636 (2013)

[65] N. Xu, C. E. Matt, E. Pomjakushina, X. Shi, R. S. Dhaka, N. C. Plumb, M. Radovic, P. K. Biswas, D. Evtushinsky, V. Zabolotnyy, J. H. Dil, K. Conder, J. Mesot, H. Ding, and M. Shi, Phys. Rev. B 90, 085148 (2014).

66 C. -H. Min, P. Lutz, S. Fiedler, B. Y. Kang, B. K. Cho, H.-D. Kim, H. Bentmann, and F. Reinert, Phys. Rev. Lett. 112, 226402 (2014).

67 J. D. Denlinger, J. W. Allen, J. -S. Kang, K. Sun, J.-W. Kim, J.H. Shim, B. I. Min, Dae-Jeong Kim, Z. Fisk, arXiv: 1312.6637 (2013).

68 N. Xu, P.K. Biswas, J.H. Dil, R.S. Dhaka, G. Landolt, S. Muff, C.E. Matt, X. Shi, N.C. Plumb, M. Radović, E. Pomjakushina, K. Conder, A. Amato, S.V. Borisenko, R. Yu, H.-M. Weng, Z. Fang, X. Dai, J. Mesot, H. Ding, M. Shi, Nat. Commun. 5, 4566 (2014).

69 V. Alexandrov, M. Dzero and P. Coleman, Phys. Rev. Lett. 111, 226403 (2013)

70 V. Alexandrov, P. Coleman, O. Erten, Phys. Rev. Lett. 114, 177202 (2015).

71 D. Kim, J. Xia and Z. Fisk, Nat. Mater. 13, 466 (2014).

72 G. Li et al., Science 346, 1208 (2014).

73 B. S. Tan et al., Science 349, 6245 (2015).

74 J. D. Denlinger et al., arXiv: 1601.07408 (2016)

75 Onur Erten, Pouyan Ghaemi, and Piers Coleman, Phys. Rev. Lett. 116, 046403 (2016)

76 B. L. Altshuler, A. A Aronov, Electron-electron interaction in disordered conductors (Elsevier, New York, ed. 1, 1985), pp. 1–159.

77 S. Thomas et al., arXiv:1307.4133 (2013).

78 J. Kirschner and R. Feder Phys. Rev. Lett. 42 1008–11 (1979)

79 R. Thiel,D. Tillmann and E. Kisker Z. Phys. B 77 1 (1989)

80 Z.-H. Pan, A.V. Fedorov, D. Gardner, Y. S. Lee, S. Chu, and T. Valla, Phys. Rev. Lett 108, 187001 (2012).

81 Z.-H. Zhu et al., Phys. Rev. Lett. 111, 216402 (2013).

82 E. Frantzeskakis et al., Phys. Rev. X 3, 041023 (2013).

83 P. Hlawenka et al., arXiv: 1502.01542 (2015).

84 J. Kirschner and R. Feder Phys. Rev. Lett. 42 1008–11 (1979)

85 R. Thiel,D. Tillmann and E. Kisker Z. Phys. B 77 1 (1989)

86 N. Sherman, Phys. Rev. 103 1601–7 (1956)

87 M. Hoesch *et al.*, Journal of Electron Spectroscopy and Related Phenomena 124   263–279 (2002)

88 J. Lobo-Checa *et al.* Phys. Rev. Lett. 104, 187602 (2010).

89 J. H. Dil, J. Phys. Condens. Matter 21, 403001 (2009).

90 U. Heinzmann and J. H. Dil, J. Phys. Condens. Matter 24, 173001 (2012).

91 K. Starke et al., Phys. Rev. B 53, R10544(R) (1996).

92 C. Jozwiak et al., Phys. Rev. B 84, 165113 (2011).

93 S. Suga et al., J. Phys. Soc. Jpn 83, 014705 (2014).

94 R. Yu, H. Weng, X. Hu, Z. Fang, and X. Dai, New J. Phys. 17, 023012 (2015).

95 J. Kim, K. Kim, C.-J. Kang, S. Kim, H. C. Choi, J.-S. Kang, J. D. Denlinger, and B. I. Min, Phys. Rev. B 90, 075131 (2014).

96 P. P. Baruselli and M. Vojta, Phys. Rev. B 90, 201106 (2014)

97 Markus Legner, Andreas Rüegg, and Manfred Sigrist, Phys. Rev. Lett. 115, 156405 (2015).

98 P. P. Baruselli and M. Vojta, Phys. Rev. Lett. 115, 156404 (2015).

99 P. Syers, Dohun Kim, M. S. Fuhrer, and J. Paglione, Phys. Rev. Lett. 114, 096601 (2015).